\newif\ifarxiv
\arxivtrue

\documentclass[preprint,12pt,times]{elsarticle}

\usepackage{amsmath,amssymb,amsthm}
\usepackage{graphicx}
\usepackage{booktabs}
\usepackage{multirow}
\usepackage{rotating}
\usepackage{algorithm}
\usepackage[noend]{algpseudocode}
\usepackage{tikz}
\usetikzlibrary{arrows.meta,positioning}
\usepackage{xcolor}
\usepackage[hidelinks]{hyperref}
\usepackage[capitalise,nameinlink]{cleveref}
\ifarxiv\else
  \usepackage{lineno}
  \linenumbers
\fi

\newcommand{\nColNinetySparsity}{89.8}

\newcommand{\nEdgeKeepMax}{1.13}

\newcommand{\nEdgeKeepMin}{0.91}

\newcommand{\nFwdMax}{0.541}

\newcommand{\nFwdMin}{0.531}

\newcommand{\nNSeeds}{3}

\newcommand{\nPhiExcessMax}{1.9}

\newcommand{\nPhiExcessMin}{1.0}

\newcommand{\nPhiNullMax}{0.524}

\newcommand{\nPhiNullMin}{0.521}

\newcommand{\nPhiNullShuffles}{20}

\newcommand{\nPhiZMin}{10}

\newcommand{\nRowsBaseFt}{98.6}

\newcommand{\nRowsBinMinusRand}{0.0}

\newcommand{\nRowsBinMinusRandSD}{0.1}

\newcommand{\nRowsCont}{89.6}

\newcommand{\nRowsDAcycStartNinetyAbs}{0.6}

\newcommand{\nRowsDFasPairMagNinetyAbs}{0.6}

\newcommand{\nRowsDFasRandNinetyAbs}{0.3}

\newcommand{\nRowsDFasRandSeventyFiveAbs}{0.2}

\newcommand{\nRowsMaxGapSeventyFive}{1.2}

\newcommand{\nRowsNoRec}{92.1}

\newcommand{\nSccFiveMax}{95}

\newcommand{\nSccFiveMin}{88}

\newcommand{\nSccTenMin}{99}

\newcommand{\nSeqBase}{95.2}

\newcommand{\nSeqBaseFt}{96.4}

\newcommand{\nSeqBinMinusRand}{\ensuremath{+}0.3}

\newcommand{\nSeqBinMinusRandSD}{1.2}

\newcommand{\nSeqColMaxNinety}{68.0}

\newcommand{\nSeqColMinNinety}{54.3}

\newcommand{\nSeqCont}{45.1}

\newcommand{\nSeqDAcycStartFiftyAbs}{2.1}
\newcommand{\nSeqDAcycStartFiftySD}{1.4}

\newcommand{\nSeqDAcycStartNinetyAbs}{10.5}
\newcommand{\nSeqDAcycStartNinetySD}{4.4}

\newcommand{\nSeqDFasPairMagFiftyAbs}{6.5}

\newcommand{\nSeqDFasPairMagNinetyAbs}{9.8}

\newcommand{\nSeqDFasRandFiftyAbs}{1.7}

\newcommand{\nSeqDFasRandNinetyAbs}{5.6}

\newcommand{\nSeqDKappaColMagNinety}{\ensuremath{-}6.6}

\newcommand{\nSeqDKappaDegNinety}{\ensuremath{-}5.1}

\newcommand{\nSeqDKappaPRNinety}{\ensuremath{-}4.8}

\newcommand{\nSeqDKappaRandNinety}{\ensuremath{+}7.0}

\newcommand{\nSeqDRewFasAcycFifty}{\ensuremath{-}0.7}

\newcommand{\nSeqDRewFasAcycNinety}{\ensuremath{+}1.3}

\newcommand{\nSeqElemNinety}{92.6}

\newcommand{\nSeqFasNinety}{76.3}

\newcommand{\nSeqFasNinetyHeight}{47}

\newcommand{\nSeqFasNinetyWidth}{14}

\newcommand{\nSeqFasRandNinety}{70.7}

\newcommand{\nSeqNoRec}{47.9}

\newcommand{\nSeqPairMagFiftyOneShotMax}{57.9}

\newcommand{\nSeqPairMagFiftyOneShotMin}{38.8}

\newcommand{\nSeqPairMagNinety}{86.2}

\newcommand{\nTieMax}{26}

\newcommand{\nTieMin}{16}

\newtheorem{theorem}{Theorem}[section]
\newtheorem{proposition}[theorem]{Proposition}
\newtheorem{lemma}[theorem]{Lemma}
\newtheorem{corollary}[theorem]{Corollary}
\theoremstyle{definition}
\newtheorem{definition}[theorem]{Definition}
\theoremstyle{remark}
\newtheorem{remark}[theorem]{Remark}

\newcommand{\down}{{\downarrow}}
\newcommand{\up}{{\uparrow}}
\newcommand{\Ideals}{\mathcal{O}}

\ifarxiv
  \makeatletter
  \def\ps@pprintTitle{%
    \let\@oddhead\@empty
    \let\@evenhead\@empty
    \def\@oddfoot{\footnotesize\itshape Revised version (supersedes arXiv:2502.16525v1)\hfill\today}%
    \let\@evenfoot\@oddfoot}
  \makeatother
\else
  \journal{Information Sciences}
\fi

\begin{document}
\emergencystretch=2em

\begin{frontmatter}

\title{Lattice-Based Pruning of Recurrent Neural Networks via Poset Modeling: A Corrected Theory and a Matched-Sparsity Evaluation}

\author{Rakesh Sengupta}
\ead{rakesh.sengupta@krea.edu.in}
\address{School of Interwoven Arts and Sciences, Krea University, Sri City 517541, Andhra Pradesh, India}

\begin{abstract}
Magnitude pruning of recurrent neural networks (RNNs) ignores how information flows between neurons. An earlier version proposed to model RNN dependencies as a partially ordered set (poset) and to retain the meet-irreducible elements of its dependency lattice. By Birkhoff's representation theorem, the meet-irreducible ideals of a finite poset are in bijection with its elements, so retaining them retains everything; the critical-set criterion used earlier selects precisely the elements comparable to all others, of which a dense recurrent layer has none. Condensing a thresholded dependency graph into strongly connected components yields a well-defined dependency order, a graded order score and an acyclic pruning rule; for Cohen--Grossberg networks and LSTMs, acyclic pruning makes local stability decompose over modules. For a 128-unit LSTM pruned at matched sparsity on row-wise and chunked sequential MNIST (three seeds, equal training budgets), the original rules remove every recurrent weight or equal random column pruning; the order score is no better than degree, PageRank or norm; trained weights show little directional bias; and imposing acyclicity from initialisation costs \nSeqDAcycStartFiftyAbs\ and \nSeqDAcycStartNinetyAbs\ percentage points at 50\% and 90\% sparsity. In the networks studied, order structure must be imposed rather than discovered.
\end{abstract}


\end{frontmatter}
\clearpage

\section{Introduction}
\label{sec:intro}

Network pruning reduces the computational cost of deep neural networks while aiming to preserve their performance \cite{blalock2020,hoefler2021,vadera2022}. Most pruning criteria score individual weights or groups of weights by magnitude or by an estimate of their effect on the loss \cite{lecun1990obd,han2015,zhu2018}. In recurrent neural networks (RNNs) the weights also define how information flows between neurons across time steps, and it is natural to ask whether this flow has a structure that a pruning criterion could exploit \cite{narang2017,wen2018iss}.

Order theory offers a precise language for such structure. A dependency relation between neurons, once it is reflexive, antisymmetric and transitive, makes the neurons a partially ordered set (poset), and the lattice of order ideals of a finite poset is a finite distributive lattice in which meet and join are intersection and union \cite{birkhoff1937,daveypriestley2002}. An earlier version of this work \cite{sengupta_v1} proposed to model RNN hidden-state dependencies, and Transformer self-attention, as posets; to construct the corresponding dependency lattices; to retain the meet-irreducible elements together with high-centrality elements; and to prune the remaining connections. Experiments on MNIST with two adjacency initialisations were reported to give sparsity--accuracy trade-offs of about 98\% accuracy at 50\% sparsity and about 87\% accuracy at 99\% sparsity.

On closer examination that proposal does not hold up in the form in which it was stated. In the present article we therefore revise it in three directions. First, we establish what the meet-irreducibility criterion actually selects. Second, we replace the dependency construction by one that is well defined for recurrent layers and that leads to criteria with genuine order-theoretic content. Third, we evaluate the original rules and the revised criteria against conventional baselines at matched sparsity, with equal training budgets and several random seeds. Our contributions are fourfold:
\begin{enumerate}
\item We show, from Birkhoff's representation theorem, that the meet-irreducible elements of the ideal lattice $\Ideals(P)$ are in bijection with the elements of $P$, so that retaining them retains every element; that the critical-set criterion of \cite{sengupta_v1}, namely that ${\down}t\setminus\{t\}$ be meet-irreducible, holds exactly when $t$ is comparable to every element of $P$; and that in a dense recurrent layer the direct-influence relation does not define a non-trivial poset at all (\cref{sec:criterion}).
\item We construct a dependency order for a weighted recurrent layer as the condensation of a thresholded dependency graph into strongly connected components, define an order score $\kappa$ from it, and introduce acyclic (cascade) pruning, which retains only connections consistent with a feedback-arc-set ordering of the neurons (\cref{sec:order,sec:acyclic}).
\item We prove that, for Cohen--Grossberg networks and for LSTMs, acyclic pruning makes the Jacobian at an equilibrium block-triangular, so that local stability can be certified module by module (\cref{sec:stability}).
\item We report a matched-sparsity evaluation on row-wise and chunked sequential MNIST that audits the original rules, compares the order score with graph-centrality scores, and measures the cost of the acyclic constraint both after training and from initialisation (\cref{sec:results}).
\end{enumerate}

The empirical conclusions are largely negative for the original programme, and we state them as such. The original rules either remove every recurrent weight or are equivalent to random column pruning; the order score does not outperform degree, PageRank or norm; and the trained LSTMs we study show only a small directional bias in their recurrent weights. What remains is a positive and measurable statement: an order can be imposed on the recurrence, it makes local stability decompose over modules, and its price in accuracy can be quantified. This article supersedes \cite{sengupta_v1}. The Transformer extension of that version is withdrawn for the reasons given in \cref{sec:limitations}.

\section{Related Works}
\label{sec:related}

Magnitude-based pruning removes the weights of smallest absolute value and fine-tunes the remaining network \cite{han2015,zhu2018}; second-order criteria such as optimal brain damage estimate the increase in loss caused by removing a weight \cite{lecun1990obd}. Structured pruning removes groups of weights so that the pruned model is smaller in a hardware-friendly sense; examples include group-sparsity regularisation \cite{wen2016} and optimisation-based weight pruning using the alternating direction method of multipliers \cite{zhang2018admm}. For recurrent networks, gradual pruning during training \cite{narang2017} and the removal of intrinsic sparse structures in LSTMs \cite{wen2018iss} have been studied. Sparse networks can also be trained from the start, either with adaptive sparse connectivity \cite{mocanu2018} or with masks found at or near initialisation \cite{frankle2019,lee2019snip}; it has been argued that the pruned architecture, rather than the inherited weights, often carries most of the benefit \cite{liu2019rethinking}. Surveys of the field emphasise that comparisons are informative only at matched sparsity, against simple baselines, and over several runs \cite{blalock2020,hoefler2021}; we follow these recommendations in \cref{sec:results}.

Dependency graphs are used in structural pruning to identify groups of parameters that must be removed together for the pruned model to remain well formed \cite{fang2023depgraph}. That notion of dependency concerns the coupling of parameters across layers. The present work concerns a different object, namely the order in which neurons influence one another through the recurrent weights. The word \emph{lattice} also has several unrelated senses in the neural-network literature: lattice algebra for single-neuron computation \cite{ritter2003}, word lattices as inputs to recurrent encoders \cite{su2017lattice}, and lattice models from statistical physics \cite{bardella2024}. Throughout this article a lattice is a partially ordered set in which every pair of elements has a least upper bound and a greatest lower bound \cite{daveypriestley2002}.

The stability theory of additive and shunting recurrent networks goes back to the Cohen--Grossberg theorem \cite{cohengrossberg1983}; contraction analysis provides conditions under which feedforward interconnections of contracting systems remain contracting \cite{lohmiller1998}. We use the elementary fact that the spectrum of a block-triangular matrix is the union of the spectra of its diagonal blocks \cite{hornjohnson2013}.

\section{Methods}
\label{sec:methods}

\subsection{Posets, ideals and lattices}
\label{sec:prelim}

Let $(P,\le)$ be a finite poset. For $t\in P$ write ${\down}t=\{u\in P: u\le t\}$ and ${\up}t=\{u\in P: u\ge t\}$. A subset $I\subseteq P$ is an \emph{order ideal} (down-set) if $u\le v\in I$ implies $u\in I$; the complement of an ideal is an up-set. The set $\Ideals(P)$ of ideals, ordered by inclusion, is a finite distributive lattice with $I\wedge J=I\cap J$ and $I\vee J=I\cup J$ \cite{birkhoff1937,daveypriestley2002}. In a finite lattice, an element $x$ other than the top is \emph{meet-irreducible} if $x=y\wedge z$ implies $x=y$ or $x=z$, equivalently if $x$ has exactly one upper cover; join-irreducibility is defined dually. An element $y$ \emph{covers} $x$ if $x<y$ and no element lies strictly between them.

\begin{lemma}[Covers in the ideal lattice]
\label{lem:covers}
Let $I\in\Ideals(P)$. The upper covers of $I$ in $\Ideals(P)$ are exactly the ideals $I\cup\{v\}$ with $v$ a minimal element of $P\setminus I$.
\end{lemma}

\begin{proof}
If $v$ is minimal in $P\setminus I$, every $u<v$ lies in $I$, so $I\cup\{v\}$ is an ideal, and it covers $I$ because it has one more element. Conversely, let $J$ cover $I$ and let $v$ be minimal in $J\setminus I$. Every $u<v$ belongs to $J$, and by minimality of $v$ it cannot belong to $J\setminus I$; hence $u\in I$. Thus $I\cup\{v\}$ is an ideal with $I\subsetneq I\cup\{v\}\subseteq J$, and since $J$ covers $I$, $J=I\cup\{v\}$. Finally $v$ is minimal in $P\setminus I$, since any $u<v$ lies in $I$.
\end{proof}

\begin{proposition}[Meet-irreducible ideals; Birkhoff]
\label{prop:birkhoff}
An ideal $I\in\Ideals(P)$ is meet-irreducible if and only if $I=P\setminus{\up}p$ for some $p\in P$. The map $p\mapsto P\setminus{\up}p$ is a bijection from $P$ onto the set of meet-irreducible elements of $\Ideals(P)$; in particular there are exactly $|P|$ of them.
\end{proposition}

\begin{proof}
By \cref{lem:covers}, $I\neq P$ is meet-irreducible if and only if the up-set $P\setminus I$ has exactly one minimal element $p$. Since $P$ is finite, every element of an up-set lies above one of its minimal elements, so this holds if and only if $P\setminus I={\up}p$, that is, $I=P\setminus{\up}p$. Distinct elements $p\neq q$ give distinct up-sets ${\up}p\neq{\up}q$ by antisymmetry, so the map is injective, and it is surjective by what has just been shown.
\end{proof}

Dually, the join-irreducible ideals are the principal ideals ${\down}p$ \cite{daveypriestley2002}. \Cref{prop:birkhoff} already implies that a pruning rule which retains \emph{the meet-irreducible elements} of $\Ideals(P)$ retains one ideal for every element of $P$, and therefore discards no element of $P$.

\subsection{What the meet-irreducibility criterion selects}
\label{sec:criterion}

The earlier version \cite{sengupta_v1} used two constructions. For RNNs, a neuron $a$ was set below a neuron $b$ when the activation of $a$ at one time step directly influences that of $b$ at the next, mutually influencing neurons were to be identified, the dependency lattice was the smallest lattice containing the resulting poset, and its meet-irreducible elements formed the critical set. For self-attention, the dependency lattice was the ideal lattice $\Ideals(P)$ of a dependency poset, and an element $t$ was declared \emph{critical} when the ideal ${\down}t\setminus\{t\}$ is meet-irreducible. In both cases connections attached to non-critical, low-centrality elements were pruned.

The first construction does not distinguish any neuron of a dense recurrent layer: every neuron directly influences every other, so the relation is symmetric, and identifying mutually influencing neurons leaves a poset with a single element. A non-trivial order requires the dependency graph to be sparsified, which we do explicitly in \cref{sec:order}. The following proposition identifies the critical set of the second construction for an arbitrary finite poset.

\begin{proposition}[Characterisation of the critical set]
\label{prop:criterion}
Let $t\in P$. The ideal ${\down}t\setminus\{t\}$ is meet-irreducible in $\Ideals(P)$ if and only if $t$ is comparable to every element of $P$.
\end{proposition}

\begin{proof}
Write $I_t={\down}t\setminus\{t\}$; it is an ideal, $t\in P\setminus I_t$, and $t$ is minimal in $P\setminus I_t$ because every $u<t$ lies in $I_t$. By the proof of \cref{prop:birkhoff}, $I_t$ is meet-irreducible if and only if $t$ is the only minimal element of $P\setminus I_t$, that is, $P\setminus I_t={\up}t$. If $t$ is comparable to every $u\in P$, each $u\notin I_t$ satisfies $u\ge t$, so $P\setminus I_t\subseteq{\up}t$; the reverse inclusion holds because no $u\ge t$ satisfies $u<t$. Conversely, if $P\setminus I_t={\up}t$, every $u$ is either in $I_t$ (so $u<t$) or in ${\up}t$ (so $u\ge t$).
\end{proof}

The characterisation is consistent with \cref{prop:birkhoff}: one always has ${\down}t\setminus\{t\}\subseteq P\setminus{\up}t$, with equality exactly when every element is either below $t$ or above it. It also shows that the critical set is degenerate in the cases that matter for recurrent networks.

\begin{corollary}[Degeneracy]
\label{cor:degenerate}
\begin{enumerate}
\item[(a)] Let $P_T=\{(i,s): i=1,\dots,n;\ s=0,\dots,T\}$ be the time-unrolled dependency poset of a dense recurrent layer, with $(i,s)\le(j,s')$ if and only if $(i,s)=(j,s')$ or $s<s'$. If $n\ge 2$, no element is critical.
\item[(b)] If $P$ is a chain, every element is critical.
\item[(c)] For $P=\{a,b,c\}$ with $a<c$ and $b<c$, the critical set is $\{c\}$.
\end{enumerate}
\end{corollary}

\begin{proof}
(a) For $n\ge 2$, $(i,s)$ and $(j,s)$ with $i\neq j$ are incomparable, so no element is comparable to all others. (b) In a chain all pairs are comparable. (c) $c$ is comparable to $a$ and $b$, whereas $a$ and $b$ are incomparable.
\end{proof}

\Cref{fig:toy} illustrates the three cases. In case (c) of \cref{cor:degenerate}, the smallest lattice containing $P$ adds a bottom element below $a$ and $b$, and its meet-irreducible elements are $a$ and $b$; the two constructions of \cite{sengupta_v1} thus select disjoint sets from the same poset, which already shows that meet-irreducibility depends on the chosen lattice rather than on a property of the network. Two further remarks place the earlier formulation in context. First, $\Ideals(P)$ is distributive, hence modular, for every finite poset $P$; modularity of the dependency lattice therefore carries no information about the network, whether its dynamics are linear or not. Second, a decomposition $x=y\wedge z$ in a lattice concerns the position of $x$ in an order; it does not imply that the activation of a neuron can be recomputed from the activations of other neurons, so meet-reducibility alone does not license the removal of a neuron's connections. The criteria introduced below are accordingly evaluated empirically rather than justified by a reconstruction argument.

\begin{figure}[t]
\centering
\begin{tikzpicture}[
  every node/.style={circle, draw=black!70, inner sep=0pt, minimum size=6.5mm, font=\small},
  crit/.style={fill=black!18, draw=black, line width=0.9pt},
  e/.style={-{Stealth[length=2mm]}, black!70},
  lab/.style={draw=none, rectangle, font=\small, minimum size=0pt}]
\foreach \s in {0,1,2}{
  \node (p\s) at (0,{\s*1.0}) {\scriptsize $1,\s$};
  \node (q\s) at (1.3,{\s*1.0}) {\scriptsize $2,\s$};
}
\foreach \s/\t in {0/1,1/2}{
  \draw[e] (p\s) -- (p\t); \draw[e] (q\s) -- (q\t);
  \draw[e] (p\s) -- (q\t); \draw[e] (q\s) -- (p\t);
}
\node[lab] at (0.65,-0.85) {(a) dense layer, unrolled};
\node[crit] (x1) at (3.6,0) {$1$};
\node[crit] (x2) at (3.6,1.0) {$2$};
\node[crit] (x3) at (3.6,2.0) {$3$};
\draw[e] (x1) -- (x2); \draw[e] (x2) -- (x3);
\node[lab] at (3.6,-0.85) {(b) chain};
\node (a) at (5.8,0) {$a$};
\node (b) at (7.1,0) {$b$};
\node[crit] (c) at (6.45,1.4) {$c$};
\draw[e] (a) -- (c); \draw[e] (b) -- (c);
\node[lab] at (6.45,-0.85) {(c) $a<c$, $b<c$};
\end{tikzpicture}
\caption{Hasse diagrams of three dependency posets; shaded elements are those for which ${\down}t\setminus\{t\}$ is meet-irreducible in $\Ideals(P)$ (\cref{prop:criterion}), matching the three cases of \cref{cor:degenerate}. (a) A dense two-neuron recurrent layer unrolled over three time steps, labelled (neuron, time): no element is selected. (b) In a chain every element is selected. (c) The poset $\{a<c,\ b<c\}$ of the toy example of \cite{sengupta_v1}: the ideal-lattice criterion selects the output $c$, whereas the meet-irreducible elements of the smallest lattice containing $P$ (obtained by adjoining a bottom element) are $a$ and $b$.}
\label{fig:toy}
\end{figure}

\subsection{A well-defined dependency order for a recurrent layer}
\label{sec:order}

Consider a recurrent layer with $H$ hidden units. For an LSTM \cite{hochreiter1997} with recurrent weight blocks $W_{hh}^{(g)}\in\mathbb{R}^{H\times H}$ for the input, forget, cell and output gates $g\in\{i,f,c,o\}$, define the \emph{dependency strength} of the influence of neuron $j$ on neuron $i$ as
\begin{equation}
S_{ij}=\sum_{g}\bigl|W_{hh}^{(g)}[i,j]\bigr|,\qquad i,j=1,\dots,H .
\label{eq:strength}
\end{equation}
In a dense layer every neuron influences every other, so the relation ``$j$ influences $i$'' is not antisymmetric, and identifying mutually influencing neurons collapses the whole layer to a single point. A non-trivial order therefore requires a sparsification of the dependency graph, which we make explicit.

\begin{definition}[Thresholded dependency graph and its condensation]
\label{def:condensation}
For $\rho\in(0,1]$ let $G_\rho$ be the directed graph on $\{1,\dots,H\}$ that contains the edge $j\to i$ ($i\neq j$) whenever $S_{ij}$ is among the largest $\rho$ fraction of off-diagonal entries of $S$. Let $C_1,\dots,C_m$ be the strongly connected components (SCCs) of $G_\rho$ \cite{tarjan1972}, and let $\bar G_\rho$ be the \emph{condensation}, the graph with an edge $C_a\to C_b$ whenever $G_\rho$ has an edge from a node of $C_a$ to a node of $C_b$ ($a\neq b$). Define $C_a\preceq C_b$ if $C_b$ is reachable from $C_a$ in $\bar G_\rho$ (including $a=b$).
\end{definition}

\begin{proposition}
\label{prop:condensation}
$(\{C_1,\dots,C_m\},\preceq)$ is a partial order.
\end{proposition}

\begin{proof}
Reachability is reflexive and transitive. The condensation of a directed graph is acyclic, since a directed cycle through two components would place all their nodes in one strongly connected component; hence $C_a\preceq C_b\preceq C_a$ implies $a=b$.
\end{proof}

The SCCs are the functional modules of the thresholded layer, and $\preceq$ orders them by the direction of influence. A natural graded score measures how much of the layer a neuron is ordered with.

\begin{definition}[Order score]
\label{def:kappa}
For a neuron $t$ in the component $C(t)$, let
\begin{equation}
\kappa(t)=\frac{\bigl|{\down}C(t)\bigr|+\bigl|{\up}C(t)\bigr|-|C(t)|-1}{H-1},
\label{eq:kappa}
\end{equation}
where ${\down}C(t)$ and ${\up}C(t)$ are taken in $\preceq$ and sizes are counted in neurons.
\end{definition}

The score $\kappa(t)\in[0,1]$ is the fraction of the other neurons that are upstream of $t$, downstream of $t$, or in the same module. When the components are singletons, $\kappa(t)=1$ exactly when $t$ is comparable to every element, which by \cref{prop:criterion} is the critical-set criterion of \cite{sengupta_v1}; $\kappa$ is its graded version. Because $\kappa$ depends on the threshold $\rho$, and because $G_\rho$ condenses into a single component once $\rho$ is moderately large (\cref{sec:structure}), we choose $\rho$ for each trained network to make $\kappa$ as informative as possible: among 25 geometrically spaced values in $[0.1\%,20\%]$ we take the one that minimises the fraction of neurons sharing the most common value of $\kappa$. Degree and PageRank \cite{page1999} are computed on the same graph $G_\rho$, so that the comparison between order and centrality is not confounded by the threshold.

\subsection{Acyclic (cascade) pruning}
\label{sec:acyclic}

Instead of searching the trained weights for an order, one can impose one. Let $\pi$ be an ordering of the neurons and let $\mathrm{pos}_\pi(i)$ be the position of neuron $i$. The connection $j\to i$ is \emph{forward} if $\mathrm{pos}_\pi(j)<\mathrm{pos}_\pi(i)$, and the \emph{forward weight fraction}
\begin{equation}
\varphi(\pi)=\frac{\sum_{i\neq j:\ \mathrm{pos}_\pi(j)<\mathrm{pos}_\pi(i)} S_{ij}}{\sum_{i\neq j} S_{ij}}
\label{eq:phi}
\end{equation}
measures how much of the recurrent weight already flows in the direction of $\pi$. Maximising $\varphi$ is the weighted feedback arc set problem; we use a variant of the greedy ordering of Eades, Lin and Smyth \cite{eades1993}, which repeatedly removes the neuron with the largest difference between outgoing and incoming weight, followed by a local search that swaps adjacent neurons while this increases $\varphi$. For a random ordering $\mathbb{E}\varphi=1/2$; the optimised ordering exceeds $1/2$ even for weights without directional structure, so we compare $\varphi$ of a trained network with the value obtained by the same procedure after shuffling the off-diagonal entries of $S$ (\cref{sec:structure}).

Given $\pi$ and a number $q$ of blocks, partition the ordered neurons into $q$ contiguous blocks $B_1,\dots,B_q$ of sizes as equal as possible, and allow the connection $j\to i$ only if $j$'s block does not come after $i$'s. The resulting mask is block-lower-triangular in the order $\pi$; connections within a block, including self-connections, are allowed. For $q=H$ the only recurrent cycles left are self-loops, and the layer becomes a cascade of single-neuron recurrences.

\begin{lemma}
\label{lem:blocksparsity}
For blocks of sizes $n_1,\dots,n_q$ with $\sum_r n_r=H$, the block-lower-triangular pattern forbids a fraction $\tfrac12\bigl(1-\sum_r n_r^2/H^2\bigr)\le\tfrac12(1-1/q)$ of the $H^2$ neuron pairs, with equality when the blocks have equal size; for $q=H$ the fraction is $(H-1)/(2H)$.
\end{lemma}

\begin{proof}
The diagonal blocks contain $\sum_r n_r^2$ pairs, and of the remaining $H^2-\sum_r n_r^2$ pairs exactly half lie above the block diagonal. By the Cauchy--Schwarz inequality $\sum_r n_r^2\ge H^2/q$, with equality for equal sizes.
\end{proof}

\Cref{alg:acyclic} summarises the procedure for a target sparsity $s$ of $W_{hh}$. By \cref{lem:blocksparsity} the pattern with $q=\lfloor 1/(1-2s)\rfloor$ blocks forbids at most a fraction $s$ of the pairs, so the target is always reachable. All four gate entries of a neuron pair are kept or removed together.

\begin{algorithm}[t]
\caption{Acyclic (cascade) pruning of an LSTM recurrent matrix}
\label{alg:acyclic}
\begin{algorithmic}[1]
\Require recurrent weights $W_{hh}\in\mathbb{R}^{4H\times H}$, target sparsity $s\in(0,1)$
\Ensure binary mask $M\in\{0,1\}^{4H\times H}$
\State compute $S$ from \eqref{eq:strength}
\State $\pi\gets$ greedy feedback-arc-set ordering of $S$ \cite{eades1993}, refined by adjacent swaps that increase $\varphi(\pi)$
\If{$s\ge (H-1)/(2H)$} $q\gets H$ \Else\ $q\gets\lfloor 1/(1-2s)\rfloor$ \EndIf
\State $A_{ij}\gets 1$ if block$_\pi(j)\le$ block$_\pi(i)$ for $q$ near-equal contiguous blocks of $\pi$, else $0$
\State keep the $\lceil(1-s)H^2\rceil$ neuron pairs $(i,j)$ with $A_{ij}=1$ and largest $S_{ij}$ (all allowed pairs if fewer)
\State $M[gH+i,j]\gets 1$ for each kept pair and every gate $g$; $0$ otherwise
\State \Return $M$
\end{algorithmic}
\end{algorithm}

\subsection{Stability under acyclic pruning}
\label{sec:stability}

The block-triangular structure has a direct dynamical consequence. Consider a Cohen--Grossberg network \cite{cohengrossberg1983}
\begin{equation}
\dot x_i=a_i(x_i)\Bigl[-b_i(x_i)+\sum_{j=1}^{H}w_{ij}f_j(x_j)+I_i\Bigr],\qquad i=1,\dots,H,
\label{eq:cg}
\end{equation}
with continuously differentiable $a_i,b_i,f_j$ and amplification functions $a_i>0$.

\begin{theorem}[Modular stability]
\label{thm:cg}
Let $B_1,\dots,B_q$ be a partition of the neurons into blocks, ordered so that $w_{ij}=0$ whenever the block of $i$ precedes the block of $j$. Let $x^*$ be an equilibrium of \eqref{eq:cg}, and for each block let
\begin{equation}
J_{rr}=D^{a}_r\bigl(-D^{b}_r+W_{rr}D^{f}_r\bigr),\qquad r=1,\dots,q,
\label{eq:jkk}
\end{equation}
where $D^{a}_r$, $D^{b}_r$ and $D^{f}_r$ are the diagonal matrices of $a_i(x^*_i)$, $b'_i(x^*_i)$ and $f'_i(x^*_i)$ for $i\in B_r$, and $W_{rr}$ is the block of $W=(w_{ij})$ on $B_r$. Then the Jacobian $J$ of \eqref{eq:cg} at $x^*$ is block-lower-triangular with diagonal blocks $J_{rr}$, and
\[
\operatorname{spec}J=\bigcup_{r=1}^{q}\operatorname{spec}J_{rr}.
\] Consequently $x^*$ is locally exponentially stable if every $J_{rr}$ is Hurwitz, and unstable if some $J_{rr}$ has an eigenvalue with positive real part.
\end{theorem}

\begin{proof}
Write the right-hand side of \eqref{eq:cg} as $a_i(x_i)g_i(x)$. At an equilibrium $a_i(x^*_i)g_i(x^*)=0$ and $a_i>0$, so $g_i(x^*)=0$. Hence
\begin{align*}
\frac{\partial(a_ig_i)}{\partial x_j}(x^*)&=a_i'(x^*_i)\,\delta_{ij}\,g_i(x^*)+a_i(x^*_i)\frac{\partial g_i}{\partial x_j}(x^*)\\
&=a_i(x^*_i)\bigl(-b_i'(x^*_i)\delta_{ij}+w_{ij}f_j'(x^*_j)\bigr),
\end{align*}
that is, $J=D^{a}(-D^{b}+WD^{f})$ with $D^{a},D^{b},D^{f}$ diagonal. Diagonal factors preserve the zero pattern of $W$ off the diagonal, so $J_{ij}=0$ whenever the block of $i$ precedes the block of $j$, and the diagonal blocks of $J$ are \eqref{eq:jkk}. The characteristic polynomial of a block-triangular matrix is the product of those of its diagonal blocks \cite{hornjohnson2013}, which gives the spectral statement. The stability statements follow from the linearisation principle for equilibria of smooth vector fields \cite{khalil2002}.
\end{proof}

Each $J_{rr}$ depends on the weights within block $B_r$ and on the equilibrium values of that block only; the upstream blocks enter solely through the location of $x^*$. Stability can therefore be checked block by block, at a cost of $\sum_r|B_r|^3$ instead of $H^3$ operations for the eigenvalue computation. Two qualifications apply. First, for $q=H$ the diagonal blocks are scalars, so the certificate becomes elementary; its value lies in the intermediate regime of a few larger modules. Second, an acyclic weight matrix cannot be symmetric unless it is diagonal, so the classical Cohen--Grossberg global stability theorem, which assumes symmetric weights \cite{cohengrossberg1983}, does not apply to acyclic networks; \cref{thm:cg} replaces it by a local, modular statement. The same local statement holds for the discrete-time LSTM.

\begin{corollary}[Acyclic LSTMs]
\label{cor:lstm}
Consider an LSTM with constant input, state $(h,c)\in\mathbb{R}^{2H}$ and update $c'=\mathbf{f}\odot c+\mathbf{u}\odot\mathbf{z}$, $h'=\mathbf{o}\odot\tanh(c')$, where each of the four gate vectors $\mathbf{u},\mathbf{f},\mathbf{z},\mathbf{o}$ depends on $h$ through the corresponding product $W_{hh}^{(g)}h$. If all four blocks $W_{hh}^{(g)}$ vanish on the pairs $(i,j)$ for which the block of $i$ precedes the block of $j$, then the Jacobian of the update at a fixed point $(h^*,c^*)$, with the state variables grouped by neuron as $(h_i,c_i)$ and ordered by blocks, is block-lower-triangular. The fixed point is locally exponentially stable if every diagonal block has spectral radius below one, and unstable if some diagonal block has spectral radius above one.
\end{corollary}

\begin{proof}
The gates of neuron $i$ depend on $h$ only through row $i$ of the matrices $W_{hh}^{(g)}$, and $c'_i$ depends on $c$ only through $c_i$; $h'_i$ depends on $c'_i$ and on the output gate of neuron $i$. Hence $\partial(h'_i,c'_i)/\partial(h_j,c_j)=0$ whenever every $W_{hh}^{(g)}[i,j]=0$ and $i\neq j$, which holds when the block of $i$ precedes the block of $j$. The spectral statement follows as in \cref{thm:cg}, and the stability statements from the discrete-time linearisation principle.
\end{proof}

\begin{remark}[Beyond local stability]
If every block of a cascade is contracting in some metric and the feedforward couplings are bounded, the whole cascade is contracting \cite{lohmiller1998}. For an acyclic network, global guarantees can therefore be composed from guarantees for the blocks, provided such guarantees are established for the blocks themselves; we do not establish them here for trained LSTMs.
\end{remark}

\begin{remark}[Height and width]
\label{rem:heightwidth}
For a pruning mask, the condensation of its neuron dependency graph is a poset whose \emph{height} (length of the longest chain) and \emph{width} (size of the largest antichain) we report in \cref{sec:results}. By Mirsky's theorem the height equals the smallest number of antichains that partition the poset \cite{mirsky1971}, and by Dilworth's theorem the width equals the smallest number of chains that cover it \cite{dilworth1950}; we compute the width as the size of a minimum chain cover through a maximum bipartite matching \cite{hopcroftkarp1973}. In an acyclic recurrence, modules in the same antichain do not feed one another, and the full trajectory of a module can be computed once the trajectories of its upstream modules are available, so the height bounds the number of sequential stages. We report these quantities descriptively and make no claim about wall-clock speed.
\end{remark}

\subsection{The pruning rules of the earlier version}
\label{sec:oldrules}

The experiments of \cite{sengupta_v1} used two rules for the recurrent matrix of an LSTM with $H$ hidden units. The \emph{binary rule} designates the first $H/2$ neurons as meet-irreducible and removes the recurrent outgoing weights (columns of $W_{hh}$) of the others. The \emph{continuous rule} draws an adjacency matrix $A$ with independent entries $A_{ij}\sim\mathcal U(0.5,2.0)$, computes the centrality $c_j=1/(\sum_iA_{ij}+\epsilon)$ of each neuron, and removes column $j$ whenever $c_j<\tau$. Neither rule reads the trained weights.

\begin{proposition}
\label{prop:oldrule}
Under the continuous rule, every column of $W_{hh}$ is removed whenever $H>2/\tau$, for every realisation of $A$.
\end{proposition}

\begin{proof}
Since $A_{ij}\ge 0.5$, $\sum_iA_{ij}\ge H/2$, so $c_j\le 2/H<\tau$ for every $j$.
\end{proof}

For the thresholds $\tau\in\{0.2,0.5,0.7\}$ of \cite{sengupta_v1} the condition holds as soon as $H>10$; in particular, at $H=128$ the continuous rule yields an LSTM without recurrent weights. The binary rule is a fixed choice of half of the columns, independent of the weights, and is therefore a particular instance of random column pruning.

\section{Results}
\label{sec:results}

\subsection{Experimental setup}
\label{sec:setup}

\paragraph{Tasks} We use MNIST \cite{lecun1998} in two sequential forms. In \emph{row-MNIST} each image is read as 28 time steps of 28 pixels. In \emph{chunked sequential MNIST} it is read in row-major order as 112 time steps of 7 pixels; this is a shorter variant of the pixel-by-pixel task of \cite{le2015irnn}, chosen because it depends strongly on recurrence while remaining inexpensive. Of the 60{,}000 training images, a fixed set of 5{,}000 is held out for validation, and results are reported on the 10{,}000 test images. Pixels are standardised with the conventional MNIST mean and standard deviation (0.1307 and 0.3081).

\paragraph{Model and training} A single-layer LSTM with $H=128$ units reads the sequence, and a linear layer maps the final hidden state to the ten classes. Training uses Adam \cite{kingma2015} with learning rate $10^{-3}$, batch size 128 and gradient-norm clipping at 1.0. Each unpruned baseline is trained for $E_0=8$ epochs. Masks are enforced at every step by zeroing both the masked weights and their gradients, so that pruned connections cannot regrow. Every configuration is run with \nNSeeds\ seeds (0, 1, 2); the seed fixes the initialisation, the order of minibatches and the random masks.

\paragraph{Sparsity} Sparsity is the fraction of zero entries in $W_{hh}\in\mathbb{R}^{4H\times H}$; input and output weights are never pruned. All methods are compared at the same sparsity of $W_{hh}$: 25\%, 50\%, 75\% and 90\% on row-MNIST, and 50\% and 90\% on chunked sequential MNIST.

\paragraph{Post-hoc pruning} Each mask is computed from the weights of the trained baseline of the same seed and applied to them. We report the test accuracy immediately after pruning (\emph{one shot}) and after $E_1=3$ epochs of fine-tuning with the mask held fixed. The equal-budget reference for fine-tuned models is the unpruned baseline trained for a further $E_1$ epochs.

\paragraph{Training from initialisation} To measure the cost of a structural constraint independently of the trained weights, a mask is applied to the initial weights of the seed and held fixed during a full run of $E_0$ epochs; the reference is the unpruned baseline, which has the same budget. Masks for this setting are either random (random neuron pairs, or a random acyclic pattern in a random neuron order) or are taken from the trained baseline and the weights rewound to their initial values \cite{frankle2019}.

\paragraph{Controls} The \emph{no-recurrence} control is an LSTM trained for $E_0$ epochs with $W_{hh}\equiv 0$; it measures the accuracy that does not require recurrent weights at all.

\paragraph{Methods} \Cref{tab:methods} lists the pruning scores. The order score, degree and PageRank are computed on the graph $G_\rho$ of \cref{def:condensation} with the automatically selected $\rho$, which lay between \nEdgeKeepMin\% and \nEdgeKeepMax\% of the off-diagonal edges for all trained networks.

\paragraph{Implementation} Experiments were implemented in PyTorch \cite{paszke2019} and NetworkX \cite{hagberg2008} and run on a single GPU. Optimiser states are re-initialised for fine-tuning. Degree is unweighted; PageRank uses the weights $S_{ij}$ and the default damping factor 0.85. Ties in the threshold selection are resolved in favour of the larger $\rho$, and ties in a column score are broken at random. Column masks remove whole columns, so their nominal 90\% sparsity is in fact $115/128=\nColNinetySparsity\%$. The validation split was not used for model selection: all configurations were fixed in advance and the test set is reported throughout. The code that produces every number, table and figure in this section, together with the raw per-seed results, is provided as supplementary material.

\begin{table}[t]
\centering
\caption{Pruning scores compared at matched sparsity of $W_{hh}$. Column methods remove the outgoing recurrent weights of a neuron (a column of all four gate blocks); pair methods remove the four gate entries of a neuron pair $(i,j)$; the unstructured method removes individual weights. Graph-based scores are computed on the thresholded dependency graph $G_\rho$.}
\label{tab:methods}
\small
\begin{tabular}{@{}lll@{}}
\toprule
Granularity & Score & Rule \\
\midrule
Column & Order score $\kappa$ & lowest $\kappa$ (\cref{def:kappa}) removed first \\
 & Degree & lowest in- plus out-degree in $G_\rho$ \\
 & PageRank & lowest PageRank of the reversed weighted $G_\rho$ (influence) \\
 & Norm & lowest $\ell_2$ norm of the column of $W_{hh}$ \\
 & Random & uniformly random columns \\
Neuron pair & Magnitude & lowest $S_{ij}$ \eqref{eq:strength} \\
 & Random & uniformly random pairs \\
Acyclic pair & FAS order & \cref{alg:acyclic} \\
 & Random order & \cref{alg:acyclic} with a random ordering $\pi$ \\
Weight & Magnitude & lowest $|w|$ \\
\bottomrule
\end{tabular}
\end{table}

\subsection{The pruning rules of the earlier version}
\label{sec:res-old}

\Cref{tab:original} reports the two rules of \cite{sengupta_v1} on the trained baselines. As predicted by \cref{prop:oldrule}, the continuous rule removes every recurrent weight at all three thresholds, and the three thresholds produce identical masks. Its accuracy after fine-tuning, \nRowsCont\% on row-MNIST and \nSeqCont\% on chunked sequential MNIST, lies below that of the no-recurrence control (\nRowsNoRec\% and \nSeqNoRec\%), which is trained without recurrent weights for the full budget. The accuracy of about 87\% reported in \cite{sengupta_v1} for this rule is therefore best understood as the accuracy of an LSTM without recurrent weights, not as evidence for a lattice-based criterion. The binary rule is indistinguishable from random column pruning at the same sparsity: the paired differences are \nRowsBinMinusRand\ $\pm$ \nRowsBinMinusRandSD\ and \nSeqBinMinusRand\ $\pm$ \nSeqBinMinusRandSD\ percentage points on the two tasks.

\begin{table}[t]
\centering
\caption{The pruning rules of \cite{sengupta_v1} applied to trained LSTMs, with references and controls. Mean $\pm$ standard deviation over \nNSeeds\ seeds. One-shot accuracy is measured immediately after pruning; final accuracy is after $E_1=3$ epochs of masked fine-tuning for pruned models, and after the stated number of epochs for references and controls.}
\label{tab:original}
\small
\resizebox{\textwidth}{!}{\begin{tabular}{@{}lrcc@{}}
\toprule
Model & $W_{hh}$ sparsity (\%) & One-shot acc.\ (\%) & Final acc.\ (\%) \\
\midrule
\multicolumn{4}{@{}l}{\textit{Row-MNIST}} \\
\quad Unpruned, $E_0$ epochs (reference) & 0.0 & --- & 98.5 $\pm$ 0.2 \\
\quad Unpruned, $E_0+E_1$ epochs (equal-budget reference) & 0.0 & --- & 98.6 $\pm$ 0.0 \\
\quad No recurrence ($W_{hh}\equiv 0$), $E_0$ epochs & 100.0 & --- & 92.1 $\pm$ 0.4 \\
\quad Continuous rule, $\tau\in\{0.2,0.5,0.7\}$ & 100.0 & 16.7 $\pm$ 2.5 & 89.6 $\pm$ 0.4 \\
\quad Binary rule (first half of neurons kept) & 50.0 & 62.4 $\pm$ 8.6 & 98.4 $\pm$ 0.2 \\
\quad Random columns, matched sparsity & 50.0 & 55.7 $\pm$ 6.0 & 98.4 $\pm$ 0.0 \\
\addlinespace
\multicolumn{4}{@{}l}{\textit{Chunked sequential MNIST}} \\
\quad Unpruned, $E_0$ epochs (reference) & 0.0 & --- & 95.2 $\pm$ 0.2 \\
\quad Unpruned, $E_0+E_1$ epochs (equal-budget reference) & 0.0 & --- & 96.4 $\pm$ 0.3 \\
\quad No recurrence ($W_{hh}\equiv 0$), $E_0$ epochs & 100.0 & --- & 47.9 $\pm$ 0.7 \\
\quad Continuous rule, $\tau\in\{0.2,0.5,0.7\}$ & 100.0 & 10.7 $\pm$ 0.1 & 45.1 $\pm$ 0.4 \\
\quad Binary rule (first half of neurons kept) & 50.0 & 14.9 $\pm$ 6.8 & 94.4 $\pm$ 0.7 \\
\quad Random columns, matched sparsity & 50.0 & 13.9 $\pm$ 6.3 & 94.1 $\pm$ 0.9 \\
\bottomrule
\end{tabular}}
\end{table}

\subsection{Order structure of trained recurrent weights}
\label{sec:structure}

\begin{figure}[t]
\centering
\includegraphics[width=\textwidth]{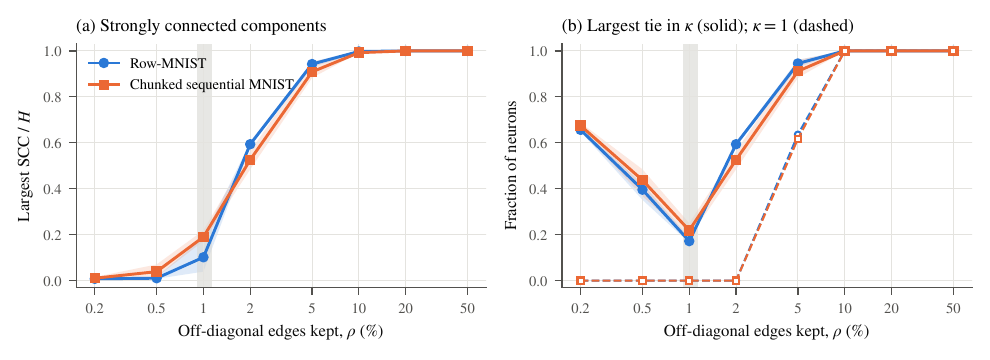}
\caption{Order structure of the recurrent weights of trained LSTMs as a function of the fraction $\rho$ of off-diagonal edges kept in $G_\rho$. (a) Size of the largest strongly connected component relative to $H$. (b) Fraction of neurons in the largest group with equal order score $\kappa$ (solid), and fraction with $\kappa=1$, the critical-set criterion (dashed). Lines are means over \nNSeeds\ seeds; bands around the solid lines span the minimum and maximum. The grey band marks the thresholds selected for the order score.}
\label{fig:scc}
\end{figure}

Before comparing criteria, we ask whether trained recurrent weights contain an order for a criterion to find. Two observations indicate that they contain little. First, the best ordering found by \cref{alg:acyclic} places between \nFwdMin\ and \nFwdMax\ of the recurrent weight mass $\sum_{i\neq j}S_{ij}$ in the forward direction, in all six trained networks. Because the optimised ordering exceeds one half even without directional structure, we compared these values with the same procedure applied to $S$ after shuffling its off-diagonal entries (\nPhiNullShuffles\ shuffles per network), which gave between \nPhiNullMin\ and \nPhiNullMax. The trained weights thus exceed the shuffled null by \nPhiExcessMin\ to \nPhiExcessMax\ percentage points of weight mass: a directional bias that is clearly detectable (at least \nPhiZMin\ null standard deviations) but small. Second, the thresholded dependency graph condenses into a single giant component as soon as a few percent of its edges are kept (\cref{fig:scc}a): with 5\% of the off-diagonal edges, the largest SCC contains between \nSccFiveMin\% and \nSccFiveMax\% of the neurons, and with 10\% it contains at least \nSccTenMin\%. The order score is informative only in a narrow window around 1\% of the edges (\cref{fig:scc}b); even at the selected threshold, between \nTieMin\% and \nTieMax\% of the neurons share the same value of $\kappa$. From 10\% of the edges upward, the criterion $\kappa=1$ of \cref{prop:criterion} marks the whole layer as critical, and at 5\% it already marks most of the layer in four of the six networks and none in the other two; at 2\% or fewer it marks none.

\subsection{Matched-sparsity comparison after training}
\label{sec:res-posthoc}

\begin{figure}[t]
\centering
\includegraphics[width=\textwidth]{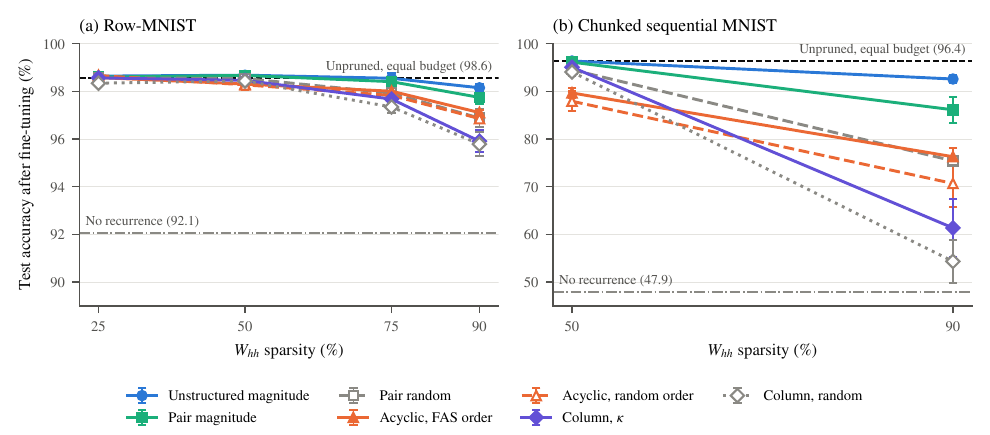}
\caption{Test accuracy after post-hoc pruning and masked fine-tuning, as a function of the sparsity of $W_{hh}$. Error bars show the standard deviation over \nNSeeds\ seeds. Degree, PageRank and norm column scores are omitted for legibility; they are reported in \cref{tab:posthoc} (on chunked sequential MNIST at 90\% sparsity they lie 4.8--6.6 points above $\kappa$). Note the different vertical scales.}
\label{fig:posthoc}
\end{figure}

\Cref{tab:posthoc} and \cref{fig:posthoc} report accuracy after masked fine-tuning. On row-MNIST every method stays, on average, within \nRowsMaxGapSeventyFive\ percentage points of the equal-budget reference (\nRowsBaseFt\%) up to 75\% sparsity, and the no-recurrence control already reaches \nRowsNoRec\%, so this task constrains the recurrent weights only weakly. Chunked sequential MNIST separates the methods: the no-recurrence control reaches only \nSeqNoRec\%, and at 90\% sparsity accuracy ranges from \nSeqElemNinety\% for unstructured magnitude pruning and \nSeqPairMagNinety\% for neuron-pair magnitude pruning, through \nSeqFasNinety\% and \nSeqFasRandNinety\% for the acyclic masks, to between \nSeqColMinNinety\% and \nSeqColMaxNinety\% for column pruning. Both the granularity of pruning and the score within a granularity matter at this sparsity; in particular, magnitude-based scores are the best at each granularity. One-shot accuracies are given in \cref{tab:oneshot} (\ref{app:oneshot}); on chunked sequential MNIST, column and acyclic masks leave the network near chance before fine-tuning at both sparsities, and all structured masks do so at 90\%, whereas neuron-pair magnitude pruning at 50\% retains between \nSeqPairMagFiftyOneShotMin\% and \nSeqPairMagFiftyOneShotMax\% accuracy.

\begin{table}[t]
\centering
\caption{Test accuracy (\%) after pruning to the stated sparsity of $W_{hh}$ and $E_1=3$ epochs of masked fine-tuning. Mean $\pm$ standard deviation over \nNSeeds\ seeds.}
\label{tab:posthoc}
\footnotesize
\setlength{\tabcolsep}{3.5pt}
\resizebox{\textwidth}{!}{\begin{tabular}{@{}llcccccc@{}}
\toprule
 & & \multicolumn{4}{c}{Row-MNIST} & \multicolumn{2}{c}{Chunked seq.\ MNIST} \\
\cmidrule(lr){3-6}\cmidrule(l){7-8}
Granularity & Score & 25\% & 50\% & 75\% & 90\% & 50\% & 90\% \\
\midrule
Unstructured & Magnitude $|w|$ & 98.6 $\pm$ 0.1 & 98.7 $\pm$ 0.1 & 98.6 $\pm$ 0.2 & 98.2 $\pm$ 0.2 & 96.4 $\pm$ 0.2 & 92.6 $\pm$ 0.7 \\
\addlinespace
Neuron pair & Magnitude $S_{ij}$ & 98.6 $\pm$ 0.1 & 98.7 $\pm$ 0.2 & 98.4 $\pm$ 0.1 & 97.8 $\pm$ 0.3 & 96.1 $\pm$ 0.3 & 86.2 $\pm$ 2.8 \\
 & Random & 98.6 $\pm$ 0.1 & 98.5 $\pm$ 0.1 & 97.9 $\pm$ 0.2 & 96.9 $\pm$ 0.4 & 94.5 $\pm$ 1.1 & 75.5 $\pm$ 0.9 \\
\addlinespace
Acyclic (cascade) & FAS order & 98.6 $\pm$ 0.2 & 98.3 $\pm$ 0.1 & 98.0 $\pm$ 0.1 & 97.1 $\pm$ 0.1 & 89.7 $\pm$ 1.0 & 76.3 $\pm$ 1.8 \\
 & Random order & 98.7 $\pm$ 0.1 & 98.3 $\pm$ 0.2 & 97.8 $\pm$ 0.1 & 96.9 $\pm$ 0.1 & 87.9 $\pm$ 2.0 & 70.7 $\pm$ 5.0 \\
\addlinespace
Column (neuron) & Order score $\kappa$ & 98.6 $\pm$ 0.1 & 98.5 $\pm$ 0.2 & 97.7 $\pm$ 0.1 & 95.9 $\pm$ 0.5 & 95.1 $\pm$ 0.5 & 61.4 $\pm$ 6.0 \\
 & Degree & 98.6 $\pm$ 0.1 & 98.5 $\pm$ 0.1 & 97.7 $\pm$ 0.0 & 96.1 $\pm$ 0.1 & 95.3 $\pm$ 0.4 & 66.5 $\pm$ 1.1 \\
 & PageRank & 98.6 $\pm$ 0.1 & 98.5 $\pm$ 0.2 & 97.9 $\pm$ 0.1 & 96.0 $\pm$ 0.3 & 95.4 $\pm$ 0.3 & 66.2 $\pm$ 8.6 \\
 & Norm $\|W_{hh}^{(:,j)}\|_2$ & 98.3 $\pm$ 0.4 & 98.4 $\pm$ 0.3 & 97.9 $\pm$ 0.1 & 95.8 $\pm$ 0.3 & 94.9 $\pm$ 0.3 & 68.0 $\pm$ 2.9 \\
 & Random & 98.3 $\pm$ 0.1 & 98.4 $\pm$ 0.0 & 97.3 $\pm$ 0.2 & 95.8 $\pm$ 0.5 & 94.1 $\pm$ 0.9 & 54.3 $\pm$ 4.5 \\
\addlinespace
Reference & Unpruned, equal budget & \multicolumn{4}{c}{98.6 $\pm$ 0.0} & \multicolumn{2}{c}{96.4 $\pm$ 0.3} \\
Control & No recurrence & \multicolumn{4}{c}{92.1 $\pm$ 0.4} & \multicolumn{2}{c}{47.9 $\pm$ 0.7} \\
\bottomrule
\end{tabular}}
\end{table}

\subsection{Order score versus centrality}
\label{sec:res-kappa}

\Cref{tab:contrasts} reports paired differences between methods within each seed. On row-MNIST the mean difference between the order score $\kappa$ and degree, PageRank or column norm is at most a few tenths of a percentage point (individual seeds differ by up to 0.8 points), and the sign is consistent across seeds only once, where $\kappa$ is worse than the column norm at 75\% sparsity. On chunked sequential MNIST at 90\% sparsity, $\kappa$ is \emph{worse} on average than degree (\nSeqDKappaDegNinety\ points), PageRank (\nSeqDKappaPRNinety) and column norm (\nSeqDKappaColMagNinety), and better than random columns (\nSeqDKappaRandNinety); these averages are dominated by one seed, and the seed-to-seed variability at this sparsity is large. One-shot differences are larger but change sign between sparsities and seeds. We conclude that the order score does not add information beyond graph centrality computed on the same graph.

\begin{sidewaystable}
\centering
\caption{Paired differences in test accuracy (percentage points) between two methods applied to the same trained network or initialisation, after fine-tuning or training unless stated otherwise. Entries are mean $\pm$ standard deviation over \nNSeeds\ seeds, with the number of seeds in which the first method was strictly better in parentheses.}
\label{tab:contrasts}
\footnotesize
\setlength{\tabcolsep}{4pt}
\resizebox{\textheight}{!}{\begin{tabular}{@{}lcccccc@{}}
\toprule
 & \multicolumn{4}{c}{Row-MNIST} & \multicolumn{2}{c}{Chunked seq.\ MNIST} \\
\cmidrule(lr){2-5}\cmidrule(l){6-7}
Contrast (percentage points) & 25\% & 50\% & 75\% & 90\% & 50\% & 90\% \\
\midrule
\multicolumn{7}{@{}l}{\textit{Order score vs.\ centrality}} \\
\quad $\kappa$ $-$ degree & 0.0 $\pm$ 0.1 (1/3) & \ensuremath{-}0.1 $\pm$ 0.2 (1/3) & 0.0 $\pm$ 0.2 (1/3) & \ensuremath{-}0.2 $\pm$ 0.5 (2/3) & \ensuremath{-}0.2 $\pm$ 0.9 (1/3) & \ensuremath{-}5.1 $\pm$ 7.1 (0/3) \\
\quad $\kappa$ $-$ PageRank & 0.0 $\pm$ 0.1 (1/3) & \ensuremath{-}0.1 $\pm$ 0.2 (1/3) & \ensuremath{-}0.2 $\pm$ 0.3 (1/3) & \ensuremath{-}0.1 $\pm$ 0.4 (1/3) & \ensuremath{-}0.3 $\pm$ 0.3 (1/3) & \ensuremath{-}4.8 $\pm$ 11.5 (1/3) \\
\quad $\kappa$ $-$ column norm & \ensuremath{+}0.3 $\pm$ 0.5 (2/3) & 0.0 $\pm$ 0.4 (2/3) & \ensuremath{-}0.2 $\pm$ 0.1 (0/3) & \ensuremath{+}0.1 $\pm$ 0.2 (2/3) & \ensuremath{+}0.2 $\pm$ 0.5 (1/3) & \ensuremath{-}6.6 $\pm$ 8.8 (0/3) \\
\quad $\kappa$ $-$ random column & \ensuremath{+}0.2 $\pm$ 0.1 (3/3) & 0.0 $\pm$ 0.3 (2/3) & \ensuremath{+}0.3 $\pm$ 0.1 (3/3) & \ensuremath{+}0.1 $\pm$ 0.6 (1/3) & \ensuremath{+}1.0 $\pm$ 0.9 (2/3) & \ensuremath{+}7.0 $\pm$ 9.6 (2/3) \\
\addlinespace
\multicolumn{7}{@{}l}{\textit{Does the order matter?}} \\
\quad Acyclic: FAS $-$ random order & \ensuremath{-}0.1 $\pm$ 0.1 (0/3) & 0.0 $\pm$ 0.3 (1/3) & \ensuremath{+}0.2 $\pm$ 0.1 (3/3) & \ensuremath{+}0.3 $\pm$ 0.1 (3/3) & \ensuremath{+}1.7 $\pm$ 2.1 (2/3) & \ensuremath{+}5.6 $\pm$ 6.7 (2/3) \\
\quad From start: rewound FAS $-$ random acyclic & --- & \ensuremath{+}0.1 $\pm$ 0.1 (2/3) & \ensuremath{+}0.1 $\pm$ 0.2 (2/3) & \ensuremath{+}0.1 $\pm$ 0.1 (2/3) & \ensuremath{-}0.7 $\pm$ 1.1 (1/3) & \ensuremath{+}1.3 $\pm$ 2.8 (2/3) \\
\addlinespace
\multicolumn{7}{@{}l}{\textit{Cost of acyclicity}} \\
\quad Post hoc: acyclic FAS $-$ pair magnitude & 0.0 $\pm$ 0.2 (1/3) & \ensuremath{-}0.4 $\pm$ 0.1 (0/3) & \ensuremath{-}0.4 $\pm$ 0.1 (0/3) & \ensuremath{-}0.6 $\pm$ 0.2 (0/3) & \ensuremath{-}6.5 $\pm$ 0.9 (0/3) & \ensuremath{-}9.8 $\pm$ 2.4 (0/3) \\
\quad Post hoc: acyclic FAS $-$ random pair & 0.0 $\pm$ 0.2 (2/3) & \ensuremath{-}0.2 $\pm$ 0.1 (0/3) & \ensuremath{+}0.1 $\pm$ 0.2 (2/3) & \ensuremath{+}0.2 $\pm$ 0.3 (2/3) & \ensuremath{-}4.9 $\pm$ 2.1 (0/3) & \ensuremath{+}0.9 $\pm$ 1.3 (2/3) \\
\quad From start: random acyclic $-$ random pair & --- & 0.0 $\pm$ 0.3 (1/3) & 0.0 $\pm$ 0.2 (2/3) & \ensuremath{-}0.6 $\pm$ 0.4 (0/3) & \ensuremath{-}2.1 $\pm$ 1.4 (0/3) & \ensuremath{-}10.5 $\pm$ 4.4 (0/3) \\
\quad From start: rewound FAS $-$ rewound pair magnitude & --- & \ensuremath{-}0.1 $\pm$ 0.1 (1/3) & 0.0 $\pm$ 0.1 (2/3) & \ensuremath{-}0.5 $\pm$ 0.2 (0/3) & \ensuremath{-}5.2 $\pm$ 0.9 (0/3) & \ensuremath{-}8.0 $\pm$ 2.7 (0/3) \\
\addlinespace
\multicolumn{7}{@{}l}{\textit{One shot (no retraining)}} \\
\quad $\kappa$ $-$ degree & \ensuremath{+}0.1 $\pm$ 2.0 (2/3) & \ensuremath{-}5.5 $\pm$ 3.4 (0/3) & \ensuremath{+}5.5 $\pm$ 4.9 (2/3) & 0.0 $\pm$ 5.5 (1/3) & \ensuremath{-}5.4 $\pm$ 5.8 (1/3) & 0.0 $\pm$ 0.1 (1/3) \\
\quad $\kappa$ $-$ random column & \ensuremath{-}0.6 $\pm$ 5.4 (2/3) & \ensuremath{-}3.0 $\pm$ 6.6 (1/3) & \ensuremath{+}4.1 $\pm$ 4.8 (3/3) & \ensuremath{+}3.0 $\pm$ 4.6 (2/3) & \ensuremath{-}4.0 $\pm$ 8.6 (1/3) & \ensuremath{-}0.2 $\pm$ 0.1 (0/3) \\
\quad Acyclic: FAS $-$ random order & \ensuremath{+}3.7 $\pm$ 5.2 (2/3) & \ensuremath{-}0.4 $\pm$ 9.0 (1/3) & \ensuremath{+}0.6 $\pm$ 3.3 (2/3) & \ensuremath{+}3.6 $\pm$ 4.0 (2/3) & \ensuremath{+}0.5 $\pm$ 0.8 (2/3) & \ensuremath{-}0.1 $\pm$ 0.1 (0/3) \\
\bottomrule
\end{tabular}}
\end{sidewaystable}

\subsection{The cost of acyclicity}
\label{sec:res-acyclic}

Acyclic pruning imposes an order rather than discovering one, and its cost can be measured directly. After training (\cref{tab:contrasts}), acyclic masks in the FAS order lose against neuron-pair magnitude pruning at every sparsity of 50\% or more, in every seed: by \nRowsDFasPairMagNinetyAbs\ points on row-MNIST at 90\% sparsity, and by \nSeqDFasPairMagFiftyAbs\ and \nSeqDFasPairMagNinetyAbs\ points on chunked sequential MNIST at 50\% and 90\%. The FAS order is better than a random order on row-MNIST at 75\% and 90\% sparsity in all three seeds, by \nRowsDFasRandSeventyFiveAbs\ and \nRowsDFasRandNinetyAbs\ points, worse at 25\% in all three seeds, and better by \nSeqDFasRandFiftyAbs\ and \nSeqDFasRandNinetyAbs\ points on chunked sequential MNIST in two of three seeds. Because the FAS ordering is chosen to keep as much weight mass as possible, part of this advantage would arise even without directional structure, so these differences do not by themselves show that the trained network contains a useful order.

\begin{figure}[t]
\centering
\includegraphics[width=\textwidth]{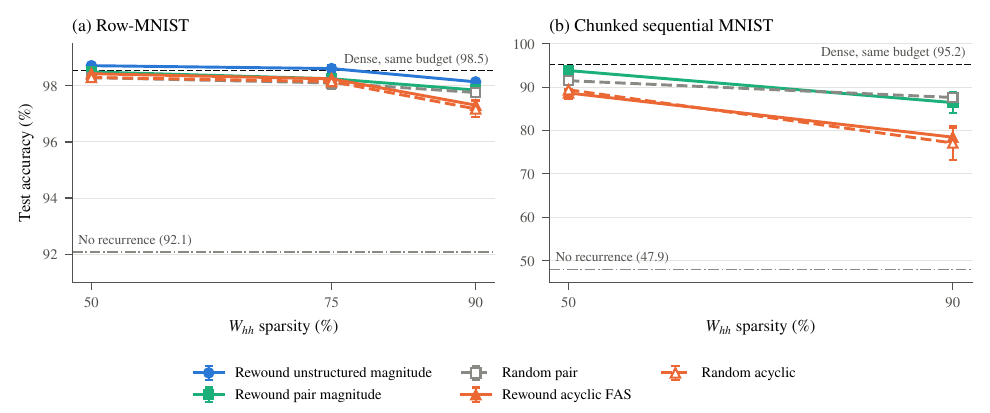}
\caption{Test accuracy when a fixed mask is trained from initialisation, as a function of the sparsity of $W_{hh}$. Error bars show the standard deviation over \nNSeeds\ seeds. Note the different vertical scales.}
\label{fig:fromstart}
\end{figure}

Training from initialisation (\cref{tab:fromstart,fig:fromstart}) isolates the cost of the constraint from the trained weights. On row-MNIST a random acyclic mask is as good as a random unconstrained mask at 50\% and 75\% sparsity and loses \nRowsDAcycStartNinetyAbs\ points at 90\%. On chunked sequential MNIST it loses \nSeqDAcycStartFiftyAbs\ $\pm$ \nSeqDAcycStartFiftySD\ points at 50\% and \nSeqDAcycStartNinetyAbs\ $\pm$ \nSeqDAcycStartNinetySD\ points at 90\% sparsity, in every seed. Rewinding an acyclic mask in the FAS order of the trained network does not reliably improve on a random acyclic mask (\nSeqDRewFasAcycFifty\ and \nSeqDRewFasAcycNinety\ points on chunked sequential MNIST). On chunked sequential MNIST at 90\% sparsity the post-hoc FAS masks have a height of about \nSeqFasNinetyHeight\ and a width of about \nSeqFasNinetyWidth\ modules (\cref{rem:heightwidth}), whereas unconstrained pair masks form a single strongly connected module apart from at most one isolated neuron.

\begin{table}[t]
\centering
\caption{Test accuracy (\%) when the mask is applied to the initial weights and held fixed for the full training budget of $E_0=8$ epochs. Rewound masks are computed from the trained baseline of the same seed. Mean $\pm$ standard deviation over \nNSeeds\ seeds. The unstructured rewound mask was evaluated on row-MNIST only.}
\label{tab:fromstart}
\footnotesize
\resizebox{\textwidth}{!}{\begin{tabular}{@{}lccccc@{}}
\toprule
 & \multicolumn{3}{c}{Row-MNIST} & \multicolumn{2}{c}{Chunked seq.\ MNIST} \\
\cmidrule(lr){2-4}\cmidrule(l){5-6}
Mask held fixed from initialisation & 50\% & 75\% & 90\% & 50\% & 90\% \\
\midrule
Random pair mask & 98.3 $\pm$ 0.1 & 98.1 $\pm$ 0.2 & 97.8 $\pm$ 0.1 & 91.5 $\pm$ 2.2 & 87.6 $\pm$ 0.6 \\
Random acyclic mask & 98.3 $\pm$ 0.1 & 98.2 $\pm$ 0.1 & 97.2 $\pm$ 0.3 & 89.4 $\pm$ 2.1 & 77.1 $\pm$ 3.8 \\
Rewound pair-magnitude mask & 98.5 $\pm$ 0.1 & 98.3 $\pm$ 0.2 & 97.9 $\pm$ 0.1 & 93.8 $\pm$ 0.4 & 86.4 $\pm$ 2.4 \\
Rewound acyclic FAS mask & 98.4 $\pm$ 0.0 & 98.2 $\pm$ 0.1 & 97.3 $\pm$ 0.2 & 88.7 $\pm$ 0.9 & 78.5 $\pm$ 2.0 \\
Rewound unstructured magnitude mask & 98.7 $\pm$ 0.1 & 98.6 $\pm$ 0.1 & 98.1 $\pm$ 0.1 & --- & --- \\
\addlinespace
Dense (same budget) & \multicolumn{3}{c}{98.5 $\pm$ 0.2} & \multicolumn{2}{c}{95.2 $\pm$ 0.2} \\
No recurrence (same budget) & \multicolumn{3}{c}{92.1 $\pm$ 0.4} & \multicolumn{2}{c}{47.9 $\pm$ 0.7} \\
\bottomrule
\end{tabular}}
\end{table}

\subsection{Numerical checks of the propositions}
\label{sec:checks}

\Cref{prop:birkhoff,prop:criterion} were checked by exhaustive enumeration of the ideal lattices of 300 random posets with 3 to 11 elements; both held in every case. \Cref{thm:cg} was checked on a Cohen--Grossberg network \eqref{eq:cg} with $H=48$, $a_i(x)=1+\tfrac12\tanh x$, $b_i(x)=x$, $f_j=\tanh$, Gaussian weights of standard deviation $1.5/\sqrt{H}$ and Gaussian inputs of standard deviation $0.5$. The neurons were ordered by the feedback-arc-set heuristic applied to $|W|$, and all connections above the block diagonal of $q=6$ equal blocks were removed (keeping 62.5\% of the absolute weight mass). At the equilibrium computed numerically (residual $8.6\times10^{-16}$), all entries of the Jacobian above the block diagonal were zero, and the eigenvalues of the Jacobian matched the union of the eigenvalues of the diagonal blocks to within $4.9\times10^{-12}$. The largest real part over the whole network, $-0.448$, coincided with the largest over the blocks.

\section{Discussion}
\label{sec:discussion}

The aim of the earlier version of this work was to use the order in which neurons influence one another as a pruning criterion. The present analysis separates three questions that were not separated there: whether the lattice-theoretic criterion is meaningful, whether trained recurrent networks contain the order that the criterion presupposes, and whether an order is useful when it is imposed.

On the first question, \cref{prop:birkhoff,prop:criterion} show that meet-irreducibility in the ideal lattice is not a selective criterion. Every element of a finite poset has its own meet-irreducible ideal, and the critical-set criterion picks out exactly the elements comparable to all others. A dense recurrent layer has no such element once it is unrolled in time, and none of its neurons is distinguished at all if the layer is quotiented by mutual influence. Any non-trivial order has to come from a sparsification of the dependency graph, and the graded score $\kappa$ of \cref{def:kappa} is the natural object to study once that sparsification is made explicit.

On the second question, the answer for the LSTMs trained here is that the recurrent weights contain little order. The forward weight fraction of the best ordering exceeds that of shuffled weights by only one to two percentage points, and the thresholded dependency graph becomes a single strongly connected component as soon as a few percent of its edges are retained. The order score is therefore informative only in a narrow range of thresholds, and even there it does not outperform degree, PageRank or column norm (\cref{sec:res-kappa}). This negative result is informative in its own right: in these networks, whatever structure an order-theoretic criterion could exploit is largely absent from the trained weights and would have to be introduced by training.

On the third question, imposing an order has a structural benefit and a measurable cost. The benefit is that, by \cref{thm:cg} and \cref{cor:lstm}, an acyclic recurrence makes the Jacobian at an equilibrium block-triangular, so that local stability can be established module by module, and global stability can be composed from contraction guarantees for the modules where such guarantees are available \cite{lohmiller1998}. The cost is in accuracy, and it depends on how much the task relies on recurrence. On row-MNIST, which an LSTM without recurrent weights already solves to \nRowsNoRec\%, the cost of acyclicity is at most a few tenths of a percentage point up to 75\% sparsity. On chunked sequential MNIST it is \nSeqDAcycStartFiftyAbs\ points at 50\% sparsity and \nSeqDAcycStartNinetyAbs\ points at 90\% when the constraint is present from initialisation, and \nSeqDFasPairMagFiftyAbs\ and \nSeqDFasPairMagNinetyAbs\ points relative to neuron-pair magnitude pruning when it is imposed on a trained network. For applications in which verifiable stability matters more than the last percentage points of accuracy, such as recurrent models of neural circuits of the Cohen--Grossberg type or recurrent controllers, this trade-off may be acceptable; for maximal accuracy at a given sparsity, unstructured or neuron-pair magnitude pruning remains preferable.

\subsection{Limitations}
\label{sec:limitations}

The evaluation concerns a single LSTM size ($H=128$), two sequential variants of MNIST, a fixed training schedule and \nNSeeds\ seeds. The seed-to-seed variability is small on row-MNIST but large on chunked sequential MNIST at 90\% sparsity, and the conclusions there rest on the sign of paired differences rather than on precise effect sizes; with three seeds, even a difference of the same sign in every seed is weak statistical evidence, and the comparisons should be repeated with more seeds and longer fine-tuning before effect sizes are quoted. The unpruned baselines on chunked sequential MNIST are not fully converged after $E_0$ epochs (\nSeqBase\% before and \nSeqBaseFt\% after the additional $E_1$ epochs); for this reason every comparison is made against a reference with the same training budget. The earlier version reported sparsities of 98.5--99.4\% for the continuous rule; the parameter set over which they were computed is not recoverable from its description, but with sparsity measured over $W_{hh}$ the rule removes 100\% of the recurrent weights (\cref{prop:oldrule}). Sparsity is measured here as the fraction of zero recurrent weights in dense tensors; we do not report wall-clock speed-ups, which would require reduced or sparse kernels. Finally, the extension to Transformer self-attention presented in \cite{sengupta_v1} is withdrawn. Attention graphs of bidirectional encoders contain cycles, including self-attention and mutual attention between tokens, so that the thresholded attention graph is not a directed acyclic graph and the dependency poset assumed there does not exist without the condensation of \cref{sec:order}; moreover that extension was evaluated on a single sentence and without a downstream task, which does not support conclusions about semantic preservation. A treatment of attention along the lines of \cref{sec:order,sec:acyclic}, evaluated on downstream tasks, is left for future work.

\section{Conclusion}
\label{sec:conclusion}

We have revisited lattice-based pruning of recurrent neural networks. In the lattice of order ideals of a dependency poset, retaining meet-irreducible elements is not selective, and the critical-set criterion of the earlier version selects only elements comparable to all others. A well-defined dependency order for a recurrent layer is obtained from the condensation of a thresholded dependency graph, and acyclic (cascade) pruning in a feedback-arc-set order makes local stability decompose over modules for Cohen--Grossberg networks and LSTMs. In a matched-sparsity evaluation with equal training budgets, the original rules reduce to removing all recurrent weights or to random column pruning, the order score does not outperform graph centrality, and the trained LSTMs carry only a small directional bias. For the networks studied, order structure therefore has to be imposed rather than discovered; doing so yields a modular stability analysis at a cost in accuracy that is small when the task depends little on recurrence and substantial when it depends on it.

\appendix
\section{One-shot accuracy}
\label{app:oneshot}

\begin{table}[h]
\centering
\caption{Test accuracy (\%) immediately after pruning the trained baseline, without fine-tuning. Mean $\pm$ standard deviation over \nNSeeds\ seeds. The reference row gives the unpruned baseline.}
\label{tab:oneshot}
\footnotesize
\setlength{\tabcolsep}{3.5pt}
\resizebox{\textwidth}{!}{\begin{tabular}{@{}llcccccc@{}}
\toprule
 & & \multicolumn{4}{c}{Row-MNIST} & \multicolumn{2}{c}{Chunked seq.\ MNIST} \\
\cmidrule(lr){3-6}\cmidrule(l){7-8}
Granularity & Score & 25\% & 50\% & 75\% & 90\% & 50\% & 90\% \\
\midrule
Unstructured & Magnitude $|w|$ & 98.5 $\pm$ 0.2 & 98.2 $\pm$ 0.4 & 90.4 $\pm$ 3.5 & 43.9 $\pm$ 2.6 & 89.1 $\pm$ 1.2 & 10.5 $\pm$ 0.5 \\
\addlinespace
Neuron pair & Magnitude $S_{ij}$ & 98.2 $\pm$ 0.2 & 95.0 $\pm$ 1.1 & 61.8 $\pm$ 2.6 & 31.1 $\pm$ 6.4 & 49.9 $\pm$ 9.9 & 9.8 $\pm$ 1.4 \\
 & Random & 95.8 $\pm$ 1.1 & 67.6 $\pm$ 4.5 & 30.9 $\pm$ 5.1 & 22.0 $\pm$ 1.6 & 12.3 $\pm$ 1.3 & 10.7 $\pm$ 0.1 \\
\addlinespace
Acyclic (cascade) & FAS order & 85.7 $\pm$ 3.0 & 43.3 $\pm$ 1.7 & 34.7 $\pm$ 2.9 & 26.7 $\pm$ 3.5 & 11.4 $\pm$ 0.1 & 10.6 $\pm$ 0.1 \\
 & Random order & 82.0 $\pm$ 4.8 & 43.6 $\pm$ 7.3 & 34.0 $\pm$ 1.2 & 23.0 $\pm$ 4.2 & 10.9 $\pm$ 0.8 & 10.7 $\pm$ 0.0 \\
\addlinespace
Column (neuron) & Order score $\kappa$ & 91.2 $\pm$ 6.3 & 52.7 $\pm$ 0.8 & 37.5 $\pm$ 4.5 & 24.0 $\pm$ 5.8 & 9.9 $\pm$ 2.7 & 10.6 $\pm$ 0.1 \\
 & Degree & 91.1 $\pm$ 4.5 & 58.1 $\pm$ 3.0 & 32.0 $\pm$ 7.1 & 24.0 $\pm$ 1.2 & 15.4 $\pm$ 3.9 & 10.6 $\pm$ 0.1 \\
 & PageRank & 93.0 $\pm$ 2.6 & 60.7 $\pm$ 2.3 & 35.6 $\pm$ 5.1 & 23.2 $\pm$ 5.5 & 10.6 $\pm$ 1.1 & 10.5 $\pm$ 0.1 \\
 & Norm $\|W_{hh}^{(:,j)}\|_2$ & 91.9 $\pm$ 3.7 & 59.0 $\pm$ 4.7 & 35.2 $\pm$ 3.8 & 22.1 $\pm$ 3.3 & 11.2 $\pm$ 1.3 & 10.6 $\pm$ 0.1 \\
 & Random & 91.8 $\pm$ 1.3 & 55.7 $\pm$ 6.0 & 33.4 $\pm$ 6.0 & 21.0 $\pm$ 1.4 & 13.9 $\pm$ 6.3 & 10.7 $\pm$ 0.1 \\
\addlinespace
Reference & Unpruned & \multicolumn{4}{c}{98.5 $\pm$ 0.2} & \multicolumn{2}{c}{95.2 $\pm$ 0.2} \\
Control & No recurrence & \multicolumn{4}{c}{92.1 $\pm$ 0.4} & \multicolumn{2}{c}{47.9 $\pm$ 0.7} \\
\bottomrule
\end{tabular}}
\end{table}

\section*{CRediT authorship contribution statement}
\textbf{Rakesh Sengupta:} Conceptualization, Methodology, Software, Formal analysis, Investigation, Visualization, Writing -- original draft, Writing -- review \& editing.

\section*{Declaration of competing interest}
The author declares that he has no known competing financial interests or personal relationships that could have appeared to influence the work reported in this paper.

\section*{Data availability}
The code that reproduces all experiments, tables and figures, and the raw per-seed results, are provided as supplementary material.

\bibliographystyle{elsarticle-num}
\bibliography{refs}

\end{document}